\documentclass[10pt,tightenlines,showpacs,aps,prl,twocolumn,amsmath,amssymb,amsfonts,floatfix]{revtex4}
\usepackage{amsxtra}
\usepackage{graphicx}
\usepackage{subfigure}
\DeclareMathOperator{\Tr}{Tr}
\begin{document}
\title{ Full Counting Statistics with Spin-sensitive
Detectors reveals Spin-singlets}
\author{Antonio \surname{Di Lorenzo}}
\author{Yuli V. \surname{Nazarov}}
\affiliation{Kavli Institute of Nanoscience,
Delft University of Technology, 2628 CJ Delft The Netherlands}
\begin{abstract}
We study the full counting statistics of electric current
to several drain terminals
with spin-dependent entrance conductances. We show
that the statistics of charge transfers can be interpreted
in terms of single electrons and spin-singlet pairs coming from
the source. If the source contains transport channels of high
transparency, a significant fraction of electrons comes
in spin-singlet pairs.
\end{abstract}
\maketitle
Spin-singlet pairs are of a fundamental interest, since they provide
a generic
example of an entangled state \cite{Einstein1935,Bohm1957}.
The entanglement presents a fundamental difference between classical and quantum
mechanics, and this has been unambiguously defined and quantified in form of theorems
involving spin-singlets \cite{Bell1964}.
The modern developments in quantum information
and manipulation have increased the interest in
experimental demonstration of entanglement.
While quantum optics presents significant experimental
advances in this direction \cite{Tittel2001},
the unambiguous experimental illustration of electron
entanglement in solid state is still to be realized.

In recent years, a significant number of
publications propose such experiment in various
solid-state nanostructures. In these proposals,
most attention is paid to production and subsequent
detection of spin-singlet pairs of electrons.
A superconductor seems to be a natural source
of spin-singlet pairs, and different schemes involving a superconductor
and normal leads have been considered: two dots  \cite{Recher2001},
two Luttinger liquids \cite{Recher2002},
two carbon nanotubes \cite{Bena2002},
or just two normal leads \cite{Lesovik2001}.
It was suggested that exchange interaction can be
used to produce singlets in a triple quantum dot device \cite{Saraga2003}
and a 2D electron gas with four point contacts \cite{Saraga2004}.
More recently, it was realized that normal leads are a source
of spin-entangled electrons~\cite{Lebedev2003}.
The current noise
was proposed to detect spin \cite{Burkard2000,Loss2000,Maitre2000,Kawabata2001} and orbital
\cite{Samuelsson2003a}
entanglement.
In Refs.~\cite{Taddei2002} the full counting statistics (FCS)
approach~\cite{Levitov1993} was used
to reveal the violation of a Clauser-Horne inequality
in multiterminal devices.
Charge current noise has been recently studied
in systems combining ferromagnets and normal
metals~\cite{Belzig2004}. FCS of spin currents
in a two-terminal device has been
addressed in ~\cite{DiLorenzo2004}.
\begin{figure}[ht]
        \subfigure[]{\includegraphics*[width=0.2\textwidth]{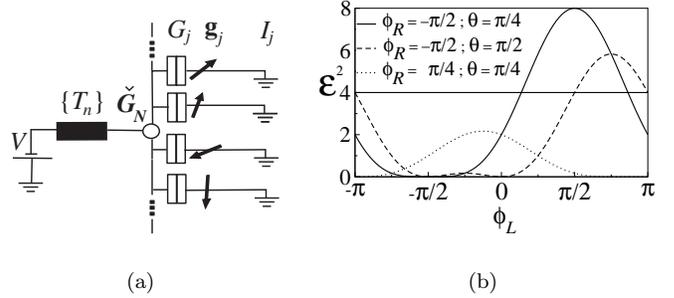}}
        \hspace{3mm}
        \subfigure[]{\includegraphics*[width=0.25\textwidth]{bellviol.eps}}
\caption{
\label{fig:setup}
(a) The setup considered. The coherent conductor
(black rectangular) is a source of electrons
being transferred to the drains: spin-sensitive
tunnel junctions. The probabilities of
one- and two-electron transfers are extracted from
measurement of the currents $I_j$ in the drains
and their correlators.
(b) The square of
Bell's parameter for the setup
described in the text
plotted versus the switching angle
$\phi_L \equiv \arccos\mathbf{n}_L{\cdot}\mathbf{n}'_L$
for several values of
$\phi_R \equiv \arccos \mathbf{n}_R{\cdot}\mathbf{n}'_R$
and
$\theta \equiv \arccos \mathbf{n}_L{\cdot}\mathbf{n}_R$
when all vectors are coplanar.
The values  of $\mathcal{E}^2$ above the horizontal line
violate Bell's inequality.
}
\end{figure}

In this Letter, we consider an almost traditional
method of spin detection that relies on spin sensitivity
of the conductance of a normal metal-ferromagnet interface~
\cite{Gijs1997}.
We demonstrate that one does not have to do anything
special to \emph{produce} spin-singlets:
they are readily present in almost any flow of
degenerate electrons, and the FCS
of currents in spin-sensitive drains reveals this
circumstance.
We do not consider any other type of entanglement than that of a spin-singlet pair.
We consider a generic coherent conductor
characterized by a set of transmission eigenvalues $T_n$.
The conductor is assumed to be short enough for
no spin-scattering taking place while an electron traverses the
conductor, so that each transport channel $n$
is spin-degenerate. The fraction of electrons coming in spin-singlet pairs
is eventually $\sum_n T^2_n /\sum_n T_n$, $2/3$ for a diffusive
conductor.

To see that a significant part of electrons comes
in spin-singlets, we concentrate on
a multi-terminal setup (Fig.~\ref{fig:setup}(a)), which consists
of a coherent conductor representing the source of electrons
and several drains, the conductances of which depends on
spin. The simplest way to achieve this is
to connect ferromagnetic
leads to a normal metallic island~\cite{Gijs1997,Brataas2000}.
Spin-sensitive conductance can also be realized with semiconductor quantum
dots in a magnetic field~\cite{Potok2002}.
In our proposal, the drains are to detect the electron propagation
via the coherent conductor. They thus should not disturb
much
the electron flow. This is ensured by the total conductance
$G_d$
of the drains being
much bigger than the conductance of the ``source".
This is easy to realize if there are
many transport channels opening to the drains, i.e.
$G_d \gg e^2/\hbar$, which we assume.
The electron spin should not change when the
electrons propagate from the ``source" to the drains, which implies
that the size of the normal island should not exceed the spin relaxation
length.
We will study zero-frequency (cross) cumulants of
electric current in the drains: the FCS of electron transfers.

The advantage of the FCS approach to quantum transport is that
it not only
gives numerical values of various cumulants
of the charge transferred, but also allows to identify elementary
independent events of transfer. 
Elementary events are defined as follows: 
the generating function (defined in Eq.~\eqref{eq:genfunc} below) 
can be presented as $S\propto \sum_n \ln{Z_{el}(n)}$, $Z_{el}(n)$ being 
a polynomial in $X_j:=\exp{i\chi_j}$.
This implies that the FCS can be interpreted as composition 
of elementary events 
characterized by the generating function $Z_{el}(n)$. 
If the coefficients of $X_1^{m_1}\cdots X_j^{m_j}\cdots$ are real and positive, they can be 
interpreted as probabilities that $m_j$ electrons were transmitted through channel $n$ to terminal $j$ 
in the course of the elementary event. 
This facilitates the interpretation
and understanding of quantum transport.
The FCS indeed reveals the many-body aspect of transport.
In principle one may have elementary events involving many particles.
Ref.~\cite{Levitov1993} shows that for two-terminal nonferromagnetic devices
elementary events involve only one electron.

It is convenient
to work with the generating function
of FCS defined in such a way that
the probability $P_{\tau}(N)$
to transfer $N$ electrons during a time interval $\tau$ reads
\begin{equation}\label{eq:genfunc}
P_\tau(N) = \int \frac{d\chi}{2\pi} \exp(S(\chi)-i\chi N)\;.
\end{equation}
For a coherent conductor biased at voltage $eV \gg k_B T$,
the
cumulant generating function (CGF)
 is given by~\cite{Levitov1993} 
\begin{equation*}
S(\chi) = \frac{eV \tau}{\pi\hbar}\sum_n \ln \left(R_n + T_n \exp(i\chi)\right)\;, (R_n \equiv 1-T_n)\;.
\end{equation*}
Interpretation of this in terms of elementary events is as follows:
In each transport channel $n$, electrons make $eV\tau/\pi\hbar$
independent attempts to traverse the conductor. An attempt is successful
with probability $T_n$. Let us generalize this to our setup assuming
at the moment that the conductances of the drains are \emph{not} spin-sensitive.
To account for electron transfers to each drain $j$, we introduce
multiple counting fields $\chi_j$. The CGF
reads
\begin{equation}
S(\{ \chi_j \}) = \frac{eV \tau}{\pi\hbar}
\sum_n \ln \left(R_n + T_n \sum_j p^{(0)}_j e^{i\chi_j}\right) \;.
\label{eq:nospin}
\end{equation}
This also allows for evident interpretation: after a successful
attempt to traverse the conductor, the electron gets to the drain $j$
with probability $p^{(0)}_j$. These probabilities are
nothing but the normalized
conductances of the drains,
$p^{(0)}_j = G_j/\sum_k G_k$, so that $\sum_j p^{(0)}_j =1$.

Now we are ready to formulate the main quantitative result
of our work. If the  conductances of the drains are spin-sensitive, the
CGF reads
\begin{align}
\nonumber
S = \frac{eV \tau}{2\pi\hbar} \sum_n
\ln&\left[R^2_n+ 2 R_n T_n \sum_j p_j e^{i\chi_j}\right.  \\
\label{eq:main}
&\left.\qquad+ T_n^2 \sum_{j,k} p_{j,k} e^{i(\chi_{j}+\chi_{k})}\right]\;.
\end{align}
The interpretation in terms of elementary events is as follows:
The electrons in each transport channel make $eV\tau/2\pi\hbar$
independent attempts to traverse the conductor. The outcomes of
each attempt are: a) with probability $R^2_n$, no electron is
transferred, b) with probability $2 R_n T_n$, one electron
traverses the conductor, c) with probability $T^2_n$, two
electrons make it. At the next stage, if one electron is
transferred, it goes to the drain $j$ with probability $p_j$. If
two electrons are transferred, the probability to have one
electron transferred to the drain $j$ and another to the drain $k$
equals $2 p_{j,k}-\delta_{jk} p_{j,j}$. If the drains are not
sensitive to spin, $p_{j,k} = p_{j}p_{k}$, and we recover 
Eq.~(\ref{eq:nospin}).
If they are, $p_{j,k} \neq p_{j}p_{k}$
in general. The concrete form of $p_{j,k}$ allows us to prove that
if two electrons are transferred, they are transferred in
\emph{spin-singlet} state. We notice that elementary processes forming
low-frequency FCS in multiterminal setups can, in principle,
involve many particles \cite{Levitov1993}. The fact that in our
specific setup an elementary event involves no more than two
particles is a result of calculation and was not assumed apriori.
Also, no expansion in the transmission values $T_n$ was adopted to limit the elementary processes 
to two-electron ones, but the formulas 
are exact analytical expressions for the setup considered. 
We show that the two-electron process is a transfer of a
spin-singlet pair, and this is the main result of our paper.

Let us give the concrete form for $p_{j,k}$. The spin-dependent
conductance of each drain can be presented as
$G_j(1 + \mathbf{g}_j {\cdot} \Hat{\boldsymbol{\sigma}})$,
$\Hat{\boldsymbol{\sigma}}$ being a vector of Pauli matrices, and
$\mathbf{g}_j$ being parallel to the magnetization direction
of the corresponding ferromagnet. The conductances for
majority (minority) spins are thus $G_j(1+|\mathbf{g}_j|)$
($G_j(1-|\mathbf{g}_j|)$). Since conductances must remain positive,
$|\mathbf{g}_j| \le 1$.
The probabilities under consideration read
\begin{subequations}
\label{elprob}
\begin{align}
\label{elprob-a}
p_j =&\ p^{(0)}_j
\frac{1-\overline{\mathbf{g}}{\cdot}\mathbf{g}_j}
{1-\overline{\mathbf{g}}^2}\;,\\
\label{elprob-b}
p_{j,k}=&\ p^{(0)}_j p^{(0)}_k \frac{1 - \mathbf{g}_j{\cdot}\mathbf{g}_k +
(\overline{\mathbf{g}}{\times}\mathbf{g}_j)
{\cdot}(\overline{\mathbf{g}}{\times}\mathbf{g}_k)}{1-\overline{\mathbf{g}}^2},
\end{align}
\end{subequations}
where we introduced a weighted quantity
$\overline{\mathbf{g}} \equiv \sum_j p^{(0)}_j \mathbf{g}_j$.
Eqs.~(\ref{eq:main},\ref{elprob}) determine the FCS in our setup
and thus present the quantitative results of our work.

Before discussing the probabilities (\ref{elprob}), their
manifestation in the (cross)cumulants of the currents, and
their relation to spin-singlets,
let us outline the derivation of Eqs.~(\ref{eq:main}), and (\ref{elprob}).
Our starting points are the Green function theory for FCS
and its circuit-theory extension to the
multi-terminal case~\cite{Nazarov2002}.
We extend this technique to spin-dependent conductances
in the spirit of Ref. \cite{Brataas2000}.
In this technique,
one works with Keldysh Green functions
that are
$4 {\times} 4$ matrices in Keldysh and spin index.

The Green functions $\check{G}_{s}$ in the source lead
and $\check{G}_j$ in
the drain leads are fixed and determined by filling factor
$f_{s}(\varepsilon), f_{j}(\varepsilon)$ and
counting field $\chi_j$ in the corresponding lead. They are
scalars in spin space and read
\[\check{G}(\chi) = e^{i\tau_z \chi/2}
\begin{pmatrix}1-2f&-2f\\-2(1-f)&2f-1
\end{pmatrix}e^{-i\tau_z \chi/2}\;,\]
$\tau_z$ being the diagonal Pauli matrix in Keldysh space.
The Green function $\check{G}_N$ in the node
is determined from the balance of (spin-dependent) matrix currents
via all the connectors~\cite{Nazarov2002}.
For the source, the matrix current is \cite{Belzig2002}
\begin{equation}
\label{source_current}
\check{I}_{s}(\{\chi_\alpha\}) =
\frac{e^2}{2\pi\hbar}
\sum_{n}
\frac{{T}_{n}\left[\check{G}_{s}
,\check{G}_N\right]}
{1+
{T}_{n} \left(\left\{
\check{G}_{s},\check{G}_N\right\}
-2\right)/4}.
\end{equation}
where $\left[...\right](\{...\})$ denote (anti)commutator of
two matrices.
As to the matrix currents through spin-sensitive
drains, they acquire spin structure.
We assume that the normal metal-ferromagnet interfaces
are tunnel junctions so that all
transmission eigenvalues $\ll 1$ : it is known that tunnel
junctions provide the best spin sensitivity~\cite{Gijs1997}.
The current can be derived with Tunneling Hamiltonian
method and reads
\begin{equation}
\check{I}_j =
\frac{G_j}{2} \left[ (1+\mathbf{g}_j {\cdot}\Hat{\boldsymbol{\sigma}})
\check{G}_j,
\check{G}_N \right]\;.
\end{equation}
This relation has been first derived in Ref.~\cite{Huertas2002} in the superconducting context and
is valid here owing to universality of matrix structure of
$\check{G}$.
Provided $\check{G}_N$ is found, the CGF 
 can be determined
from the relation $\partial S /\partial \chi_j =
(\tau/8 e^2) \int d\varepsilon \Tr\{\tau_z \check{I}_j\}$.
To determine $\check{G}_N$, one generally has to solve the
current balance equation $\check{I}_s +\sum_j \check{I}_j=0$, 
with the constraints $\check{G}_N^2 = 1$ and $\Tr{\check{G}_N}=0$.
However, in the general case, the solution is complicated
by
multiple electron trips from the drains to the source
and back. We do not wish to account for this, since in our
setup the drains are merely detectors and are not supposed to
perturb the electron flow. So we will solve this equation only
in the corresponding limiting case $ G_d \gg G_s$.
This means that $\check{G}_N$ is to be determined from the balance
of the drain currents only, $\sum_j \check{I}_j=0$. The source
matrix current is found by substitution of the solution into
Eq.~(\ref{source_current}). We consider only the shot noise
limit $eV \gg k_B T$. In this case, the contribution
to FCS comes from the energy strip of width $eV$ where $f_s =1, f_j=0$.
From current conservation,
one proves that in this limit
\[
S = \frac{eV}{2} \sum_n \Tr \ln \left\{ 1+ \frac{T_n}{4}
\left(\{ \check{G}_s, \check{G}_N \}-2\right)\right\} \;.
\]
We have that the solution $\check{G}_N$
reads 
\[
\check{G}_N = 
\begin{pmatrix}
1&0\\
\beta_0+\boldsymbol{\beta}\cdot \Hat{\boldsymbol{\sigma}}&-1
\end{pmatrix}\;,
\]
with 
$\beta_0=-2 (p_\chi-\overline{\mathbf{g}}\cdot\mathbf{g}_\chi)
/(1-\overline{\mathbf{g}}^2)$, 
$\boldsymbol{\beta}= 
-2 (\mathbf{g}_\chi-p_\chi \overline{\mathbf{g}}+
\overline{\mathbf{g}}\times(\overline{\mathbf{g}}\times\mathbf{g}_\chi))
/(1-\overline{\mathbf{g}}^2)$, $p_\chi\equiv \sum_j (G_j/G_d) e^{i\chi_j}$, and
$\mathbf{g}_\chi\equiv \sum_j (G_j/G_d) e^{i\chi_j} \mathbf{g}_j$. 
Substituting the concrete expressions for the Green functions,
we arrive at Eq.~(\ref{eq:main}), with probabilities given by
(\ref{elprob}).

Now we are in the position to discuss and interpret
the probabilities (\ref{elprob}). Let us first consider
the case $\overline{\mathbf{g}}=0$. Although it
is 
not the most general case, the
conductances of the drains can be 
tuned to achieve this.
The expressions for probabilities are 
simpler in this
case:
\begin{equation}
\label{elprob1}
p_j = p^{(0)}_j;\; p_{j,k}= p^{(0)}_j p^{(0)}_k
(1 - \mathbf{g}_j{\cdot}\mathbf{g}_k )\;.
\end{equation}
Thus,
the one-electron probability to get into a certain
drain does not depend on all other drains, except that
it is determined by the normalized conductance of the drain.
It is not sensitive to electron spin either.
In contrast to this, the two-electron probability
does depend on spin. The concrete expression for two-electron
probability can be re-derived
if one starts with the two-particle density matrix of the spin-singlet
state
\begin{equation}
        \Hat{\rho}_{sing} = \frac{1}{4} \left(
\Hat{1} - \Hat{\boldsymbol{\sigma}}_1 {\cdot}
\Hat{\boldsymbol{\sigma}}_2\right)\;,
\end{equation}
$1, 2$ numbering the particles.
For one particle, the probability to tunnel
to a certain drain is proportional to
$\Tr{\left\{{G}_j(1+\mathbf{g}_j{\cdot}\Hat{\boldsymbol{\sigma}})
\Hat{\rho}\right\}}$.
Consequently,
the probability for two particles to tunnel
to the drains $j$ and $k$ is proportional
to $\Tr{\left\{
G_j G_k (1+\mathbf{g}_j{\cdot}\Hat{\boldsymbol{\sigma}}_1)
(1+\mathbf{g}_k{\cdot}\Hat{\boldsymbol{\sigma}}_2)
\Hat{\rho}\right\}} + ( 1 \leftrightarrow 2)$.
Using the spin-singlet density matrix, we recover
Eq.~(\ref{elprob1}). This not only
means that the electrons come with opposite spin, but also the
probabilities distinguish between spin-singlet and
a component of the triplet state with opposite spin.

Let us put these probabilities in the context of general
discussion of the relations between locality and quantum
entanglement that provide the initial fascination with
the subject \cite{Einstein1935}.
Let us assign classical observers, $Alice$ and $Bob$
to  two of the drains.
Let us also disregard
the current fluctuations of the source
and just let it pass a fixed number of electrons.
The observers can change the
direction of $\mathbf{g}_{A,B}$ in their own drains.
If only one-electron processes occur, the 
readings of $Alice$ and $Bob$ would be 
uncorrelated. One can
interpret this as locality: the counted electrons 
are independent, and an electron
counted by $Alice$ would never get to $Bob$ passing
information about direction of $\mathbf{g}_A$.
However, if spin-singlets are coming, the readings do
correlate by virtue of Eq.~(\ref{elprob1}).
If $Bob$ knows
the readings of $Alice$,
he can compare it with his own observations and figure
out the direction of her $\mathbf{g}_{A}$\ \cite{Dieks1982}.

In the general case, $\overline{\mathbf{g}}\ne 0$,
the probabilities are less
straightforward
to interpret.
The one-electron probability for a given
drain does depend on the $\mathbf{g}_j$ of the other drains
by means of $\overline{\mathbf{g}}$. The reason for
this is transparent: there is \emph{spin accumulation}
in the node~\cite{Gijs1997,Brataas2000}.
The $-\overline{\mathbf{g}}$
is the
average polarization
of electrons in the node. In fact, the one-electron
probabilities can be re-derived assuming that one-particle
density matrix reads $\Hat{\rho} =(\Hat{1}-\overline{\mathbf{g}}
{\cdot}\Hat{\boldsymbol{\sigma}})/2$. The two-particle density
matrix reads
\begin{equation}\label{eq:spinaccro}
\Hat{\rho} = \frac{1}{4} \left[
\Hat{1}-(1- \overline{\mathbf{g}}^2) \Hat{\boldsymbol{\sigma}}_1 {\cdot}
\Hat{\boldsymbol{\sigma}}_2 -
(\overline{\mathbf{g}}{\cdot}\Hat{\boldsymbol{\sigma}}_1)
( \overline{\mathbf{g}}{\cdot}\Hat{\boldsymbol{\sigma}}_2)
\right],
\end{equation}
This density matrix is the mixture of the singlet density matrix and
of the
$(\Hat{\boldsymbol{\sigma}}_1+\Hat{\boldsymbol{\sigma}}_2) {\cdot} \overline{\mathbf{g}}=0$
component of the triplet density matrix,
with respective weights $1- \overline{\mathbf{g}}^2/2$ and $\overline{\mathbf{g}}^2/2$.
We stress that Eq.~\eqref{eq:spinaccro} gives the density matrix of electrons
that go to drains, and not the one of electrons coming
from the source: They still come in spin-singlet pairs.
Spin accumulation thus distorts this matrix, both
for single electrons and electron pairs.

The elementary probabilities $p_j, p_{j,k}$
can be readily extracted from 
the measurement
of 
average currents $I_j$ and
low-frequency noise (correlations) $S_{jk}$ in the drains,
since
\begin{subequations}\label{currcorr}
\vspace{-5mm}
\begin{align}
\label{currcorr-a}
 I_{j} =& I  p_j \;,\\
\label{currcorr-b}
S_{jk} =&\
2eI \left[ \Theta
(p_{j,k}-2p_j p_k) +
p_j\delta_{j,k} \right] \;,
\end{align}
\end{subequations}
\mbox{$I$ being the current in the source, and $\Theta \equiv \sum_n
T^2_n/\sum_n T_n$} giving the fraction of electrons coming in
spin-singlet pairs. $\Theta$ 
is related to 
the Fano
factor $F$ (ratio of noise to $2eI$) of the source, 
$\Theta = 1 - F$. We notice that, in contrast to optics, the
measurement is not time resolved, since we access FCS at low
frequency, at times much longer than time intervals between
electron transfers.

The simplest illustration is a ballistic quantum
point contact ($T_n =1$) as source and two drains 1, 2 that
can only accept electrons with spin up (1) and down (2).
This implies that \emph{all} electrons come
in spin-singlet pairs and,
since $g^{z}_1 = - g^{z}_2 = 1$, $p_1=p_2=1/2$, $p_{11} =
p_{22} =0$, $p_{12}=1/2$. Therefore the same amount of
electrons go to the drains 1 and 2, and there is an absolute
correlation of the currents in the drains,
$S_{11}=S_{22}=S_{12}=0$!

We demonstrate that our setup can be used
to illustrate the violation of a
Bell-Clauser-Horne-Shimony-Holt inequality\ \cite{Bell1964}.
We consider four drains. The drains $1, 2$ and $3, 4$
are pairwise antiparallel, i.e.
$\mathbf{g}_1=-\mathbf{g}_2=\mathbf{g}_L$,
$ \mathbf{g}_3=-\mathbf{g}_4=\mathbf{g}_R$.
For simplicity, $G_1=G_2$ and $G_3=G_4$.

This is a close analogue of the optical experiments
\cite{Tittel2001}.
Each pair of drains forms a ``spin detector": e.g.~the
current through drain 1 (2)
measures the number of electrons coming
with spin up (down) with respect to $\mathbf{g}_L$.
The $|\mathbf{g}_{L,R}|$ turn out to be the detectors'
efficiencies.
The measurements are performed with each polarization
taking two directions, $\mathbf{n}^{\phantom{'}}_{L,R}, \mathbf{n}'_{L,R}$
($\mathbf{n}^{2}_{L,R}={\mathbf{n}'}^2_{\!\!L,R}=1$).
We shall discard the events where
both electrons of a singlet
go to the same detector by normalizing the probabilities
to go to different detectors. For instance,
the probability to measure spin up in the left and
right detectors reads $P_{++} = p_{1,3}/(p_{1,3} + p_{1,4}+
p_{2,3} + p_{2,4})$.

The Bell parameter is defined
as ${\mathcal E} \equiv \lvert E(\mathbf{n}_L,\mathbf{n}_R) +
E(\mathbf{n}'_L,\mathbf{n}_R)  +E(\mathbf{n}_L,\mathbf{n}'_R) -
E(\mathbf{n}'_L,\mathbf{n}'_R) \rvert$
where the correlator is given by
$E(\mathbf{n},\mathbf{n'}) =
P_{++}+P_{--}-P_{+-}-P_{-+}$.
From
Eq.~(\ref{elprob1}) we obtain that
$E =- \mathbf{g}_L {\cdot} \mathbf{g}_R $. The
Bell parameter is proportional to efficiencies of both
detectors, $\mathcal{E} =
\lvert \mathbf{g}_L\rvert \lvert \mathbf{g}_R\rvert \mathcal{E}_0$,
where
$
\mathcal{E}_0
=\left| {\mathbf{n}}_L{\cdot} {\mathbf{n}}_R +
{\mathbf{n}}'_L{\cdot} {\mathbf{n}}_R +
{\mathbf{n}}_L{\cdot} {\mathbf{n}}'_R -
{\mathbf{n}}'_L{\cdot} {\mathbf{n}}'_R\right|
$
is
the
expression
for fully efficient detectors
from the work of Bell.
Since the maximum value of $\mathcal{E}_0$
is $2\sqrt{2}$, Bell's inequality
$\mathcal{E}\le 2$
is violated at certain configurations of $\mathbf{n}$
provided the polarization exceeds
$2^{-1/4}\simeq 84\%$, in agreement with~\cite{Kawabata2001}. 
We plot $\mathcal{E}$ in Fig.~\ref{fig:setup}(b).

To conclude, we have shown that the low-frequency
FCS of a coherent conductor can be interpreted
in terms of single-electron and spin-singlet pairs transfers.
This can be revealed and quantified by using spin-sensitive
drains.

We acknowledge the financial support provided through the European
Community's Research Training Networks Programme under contract
HPRN-CT-2002-00302, Spintronics.

\end{document}